\documentclass{article}

\usepackage{arxiv}
\usepackage{booktabs}
\usepackage{tabularx}
\usepackage[utf8]{inputenc} 
\usepackage[T1]{fontenc}    
\usepackage{hyperref}       
\usepackage{url}            
\usepackage{booktabs}       
\usepackage{amsfonts}       
\usepackage{nicefrac}       
\usepackage{microtype}      
\usepackage{lipsum}		
\usepackage{graphicx}
\usepackage{tikz}
\usepackage{subcaption}
\usepackage[authoryear]{natbib}
\usepackage{doi}
\usepackage{amsmath}

\title{Revisiting Gossip Protocols: A Vision for Emergent Coordination in Agentic Multi-Agent Systems}

\author{ \href{https://orcid.org/0000-0001-9051-1370}{\includegraphics[scale=0.06]{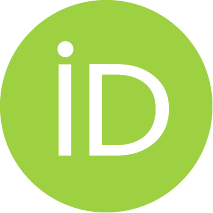}\hspace{1mm}Mansura habiba} \\
	Principal Platform Architect - Agentic AI\\
	IBM Software\\
	Dublin, Ireland \\
	\texttt{mansura.habiba@gmail.com} \\
	\And
	{\hspace{1mm}Nafiul I. Khan} \\
	Department of Computer Technology\\
	City Polletechnique College\\
	Khulna, Bangladesh \\
	\texttt{earthkhan01@gmail.com} \\
}

\renewcommand{\shorttitle}{The Gossip Layer: Enhancing Agent Communication in Distributed AI}

\hypersetup{
pdftitle={A template for the arxiv style},
pdfsubject={q-bio.NC, q-bio.QM},
pdfauthor={David S.~Hippocampus, Elias D.~Striatum},
pdfkeywords={First keyword, Second keyword, More},
}

\begin{document}
\maketitle

\begin{abstract}
As agentic platforms scale, agents are evolving beyond static roles and toolchains, creating a pressing need for flexible, decentralized coordination. Today’s structured protocols for multi-agent communication (e.g., direct agent-to-agent messaging) excel at reliability. However, we still have to solve emergent, swarm-like intelligence where distributed agents grow an organic intelligence among them through continuous learning and communication. In this paper, we revisit a well-known protocol, Gossip protocols, which is widely valued in distributed systems for its decentralized, fault-tolerant properties, offering a missing layer of adaptive, context-sharing communication to fill this gap. However, Gossip is not without challenges: it introduces issues of semantic relevance, temporal staleness, and limited action consistency, especially in high-stakes or fast-changing environments. We explore how gossip supports scalable, adaptive, and context-rich information diffusion across agents—filling gaps that structured protocols alone cannot efficiently address. At the same time, we confront key limitations: semantic filtering, temporal staleness, and trustworthiness in peer-to-peer exchange. We map out where Gossip could reshape the assumptions of agent-to-agent design, identify open questions around trust, context decay, and coordination under uncertainty, and suggest why this line of thinking deserves more space in the research agenda. Gossip may not be a silver bullet—but ignoring it risks missing a path toward more resilient, emergent forms of agentic intelligence. Rather than propose a complete framework, this work charts a research agenda for rethinking agentic communication around Gossip as a complementary substrate. We identify foundational questions around coordination, trust, and knowledge decay, and suggest why embracing Gossip is essential for the future of resilient, self-organizing multi-agent systems.
\end{abstract}


\section{Introduction}
As the agentic paradigm continues to mature and become part of a wide array of real-world applications, its adoption is also accelerating. Presently, agents are primarily employed to automate complex but well-defined processes. However, this static arrangement is poised to evolve. With continued development, agents are expected to become more autonomous—not only in their decision-making capabilities but also in defining and optimizing their orchestration logic. As this shift unfolds, it is crucial to recognize that agents will require access to richer, emergent knowledge and more robust mechanisms for communicating with each other. Future agents will not simply execute pre-defined steps—they will adapt their behavior based on real-time environmental stimuli, shared insights from other agents, and historical learning. They are no longer limited to executing isolated tasks or deterministic workflows. Agents are beginning to operate as adaptive entities—learning from local observations, evolving goals, and the behavior of peers. In this emerging paradigm, agents must not only access tools and data but also share context, infer intent, and collectively adapt in open-ended environments.

This evolution demands a new kind of communication substrate—one that goes beyond pre-defined interfaces and direct messaging. Today's agent communication protocols—such as Model Context Protocol (MCP) \citep{mcp-gitHub, mcp} and Agent-to-Agent (A2A)\cite{a2a}—support structured, secure task delegation. But they are fundamentally centralized, deterministic, and inadequate for the reflexive, adaptive behaviors that will define future agentic systems.

We envision a complementary class of agent communication protocols, inspired by the Gossip \cite{gossip}  mechanisms long used in distributed systems, which can offer a radically different design: decentralized, fault-tolerant, and emergent. We argue that gossip should become a first-class citizen in agent communication—not as a replacement for structured protocols, but as a foundational augmentation.

This paper proposes a foundational vision for designing such protocols—not as retrofits to existing standards, but as first-class constructs. We draw from decades of distributed computing, biological swarms, and emergent AI behavior to explore the properties a gossip-inspired communication layer could bring to agentic AI systems. Rather than defining a rigid specification, we outline the conceptual pillars of this paradigm: semantic propagation, intent gossiping, adaptive redundancy, and asynchronous consensus.

The main contribution of this paper includes:
\begin{itemize}
  \item Analyze the requirement of a multi-agent system in supporting emergent behavior,
  \item Outline the core properties of gossip protocols relevant to agentic coordination,
  \item Define a future research agenda, including trust models, semantic filtering, and integration with learning systems.
\end{itemize}

We believe this vision marks a necessary evolution in agent protocol design—toward systems that not only talk, but collectively learn, adapt, and evolve.

\section{Current Agent Communication Protocols}

Modern agentic systems rely increasingly on structured communication protocols to ensure reliable, secure, and interpretable interactions among intelligent agents and tools. Two prominent efforts in this domain, Anthropic's \textbf{Model Context Protocol (MCP)} and Google's \textbf{Agent-to-Agent (A2A)} protocol, represent the state of practice in enabling modular agent coordination, tool invocation, and cross-context task delegation.

While these protocols provide essential foundations for structured orchestration, they also reflect design assumptions: stable topologies, pre-defined roles, and explicit, deterministic messaging. As agentic systems scale toward distributed, reflexive, and open-ended deployments, new coordination requirements are emerging—ones that structured protocols alone may not fully address. 

\subsection{MCP (Model Context Protocol)}

MCP is an open standard designed to connect AI assistants to external tools and data sources in a secure, uniform way \cite{mcp}. Essentially, MCP enables agents (such as large language model-based assistants) to fetch data or invoke tools through a standardized API and context formats, rather than primarily serving as an agent-to-agent protocol, but rather as an agent-to-environment interface. For example, using MCP, an AI agent can query a company database or run a computation in a sandbox, with the protocol handling authentication and data formatting. Structured nature: MCP messages have a defined schema (JSON-RPC over HTTP in current implementations) and each interaction is a direct request-response: the agent asks for something, the tool responds
\cite{developers.googleblog.com}. This is structured communication — every message has a clear purpose, format, and expected reply. When considering multi-agent systems, MCP can be utilized for an agent to request Information from another agent that exposes a tool interface. However, MCP by itself doesn't define how two peer agents spontaneously talk; it's more about giving agents access to contextual data and operations. One could imagine a scenario where each agent runs an MCP server providing some services (like access to its local observations), and other agents connect via MCP queries. In effect, that would be a query-based interaction model.

However, MCP assumes a central initiator for all orchestration decisions. Tools and agents must be registered ahead of time, discovery is static, and communication relies on direct, pre-declared invocation patterns. While MCP addresses many foundational needs for secure and modular agent behavior, its design is bounded by assumptions of centralized orchestration and preconfigured execution flows. As agentic ecosystems evolve toward more decentralized and reflexive paradigms, several limitations become apparent:

\begin{itemize}
    \item \textbf{No runtime peer discovery.} All agents or tools must be explicitly registered in advance. MCP lacks a native mechanism for dynamically advertising, discovering, or querying new peers or capabilities, which constrains flexibility in open or evolving networks.
    
    \item \textbf{Limited fault tolerance:} MCP provides no built-in mechanism for agent failover, replication, or consensus. Recovery from failure relies on external retry logic or manual orchestration, making the system brittle under dynamic or lossy conditions.
    \item \textbf{No distributed coordination primitives:}. Tasks are routed deterministically by a central orchestrator. There is no notion of leader election, quorum consensus, or distributed role assignment as found in protocols like RAFT or gossip-based election.
    \item \textbf{Lack of emergent delegation:}Although agents can dynamically invoke tools, all decisions must be initiated by the orchestrating entity. There is no framework for bottom-up task claiming, bidding, or peer negotiation.
    \item \textbf{No shared memory model:} While MCP supports session-level memory, it lacks constructs for collaborative state convergence (e.g., CRDTs or blackboard architectures). Multi-agent collaboration must be engineered manually via shared databases or custom APIs.
    \item \textbf{Swarm-style organization:} MCP does not support collective behavior without central control. Patterns such as decentralized delegation, emergent clustering, or reflexive adaptation are outside its specification.
\end{itemize}

These limitations do not diminish MCP's utility in structured environments. Still, they highlight a growing architectural gap: the need for communication substrates that support \emph{emergence, fault tolerance, and adaptive behavior} in decentralized agent collectives. As we explore these gaps further, it becomes natural to consider alternative paradigms—such as gossip-based communication—not as replacements, but as candidate extensions to the agentic protocol landscape.

\subsection{Agent-to-Agent Protocol (A2A)}

The Agent-to-Agent (A2A) protocol, proposed by Google~\cite{developers.googleblog.com}, defines a schema for inter-agent communication using web-native constructs such as HTTP, JSON, and Server-Sent Events (SSE). A2A formalizes structured task-based messaging between agents, allowing one agent to delegate or request a task from another, complete with metadata, roles, and task lifecycle management.

Each exchange enforces a client-responder structure, mirroring remote procedure call (RPC) semantics across autonomous entities. A2A supports heterogeneous agents (cross-vendor or cross-runtime), multimodal payloads, and secure provenance through OAuth2-compatible mechanisms.

A2A is particularly well-suited for pairwise agent collaboration, but—like MCP—it presumes pre-defined roles, static discovery, and explicit coordination. It does not provide a framework for spontaneous collaboration, distributed consensus, or emergent task sharing in dynamic environments.

Despite recent advances in agent communication protocols, effective orchestration in decentralized multi-agent systems remains a challenge. Google's Agent-to-Agent (A2A) Protocol offers a standardized schema for inter-agent messaging but does not sufficiently address the orchestration demands that emerge in real-world distributed environments. Through a focused evaluation of A2A's design, we identify critical coordination gaps and explore how decentralized Gossip protocols can complement these shortcomings.
\begin{itemize}
    \item \textbf{Decentralized Coordination:}At the core of distributed agent systems lies the need for decentralized coordination—agents must not only exchange messages but also independently determine how to route tasks, adjust workflows, or synchronize actions. A2A excels at defining the format and structure for message exchange, but it stops short of offering semantics or guidance on coordination logic. In contrast, Gossip protocols naturally support local propagation of context and task offers, enabling coordination to emerge from repeated, lightweight exchanges among peers.
    \item \textbf{Fault-Tolerant, Leaderless Execution:} Another gap in A2A is its laLackf support for fault-tolerant, leaderless execution. There is no built-in mechanism for electing leaders, establishing consensus, or recovering from agent failures. This makes the protocol fragile in highly dynamic or unreliable environments. Gossip-based systems, on the other hand, are inherently robust in such scenarios. By diffusing status information and fallback strategies throughout the network, they allow agents to react and reorganize locally in the event of leader failure or communication dropouts.
    \item \textbf{Emergent Delegation and Load Distribution:} A2A assumes agents already know whom to contact—a reasonable assumption in controlled deployments, but a limiting one in open-ended systems where roles may change, or load may need to be redistributed dynamically. It lacks support for broadcasting task availability or enabling agents to bid or claim tasks on the fly. Gossip protocols offer a compelling alternative here by allowing agents to share load signals or task offers locally, supporting emergent delegation based on proximity, capacity, or interest.
    \item \textbf{Peer Discovery at Runtime:} While A2A does provide a way for agents to advertise metadata, it does not offer a fully dynamic peer discovery mechanism. Agents must rely on pre-registered services or static lookup. This limits their ability to adapt to changes in the network topology. Gossip protocols, in contrast, enable live peer discovery through continuous, lightweight information exchanges, ensuring agents always have an up-to-date view of their local network.
    \item \textbf{Distributed State and Consensus:} The absence of state replication or consensus mechanisms is another limitation of A2A. There is no notion of shared state or anti-entropy synchronization, which hinders consistency in distributed decision-making. Gossip protocols can be layered with CRDTs (Conflict-Free Replicated Data Types) or similar techniques to enable eventual consistency of shared state across agents, allowing for robust collaboration without centralized control.
    \item \textbf{Shared Memory and Knowledge Propagation:} A2A does not support a blackboard-style architecture where agents can read from and write to a shared memory space. This limits their ability to collaborate in loosely coupled ways, where knowledge contributions may accumulate and influence subsequent decisions. Gossip-based techniques can simulate such shared memory through repeated Information spreading and convergence, effectively allowing agents to coalesce around shared facts or partial knowledge.
\end{itemize}

In conclusion, while the A2A protocol lays the foundation for interoperable communication among agents, it lacks orchestration primitives necessary for decentralized, adaptive, and resilient agent ecosystems. Gossip protocols provide a promising augmentation, introducing mechanisms for emergent behavior, dynamic coordination, and fault tolerance. Future work should explore hybrid models where formal message-passing (as enabled by A2A) is paired with informal Gossip layers to achieve more prosperous and more resilient agentic architectures.

\subsection{Emerging Needs in Distributed Agentic Systems}

As Agentic AI shifts from centralized pipelines to decentralized, self-organizing networks of autonomous agents, structured communication models face new pressures. Use cases such as distributed monitoring, resilient swarms, emergent planning, and asynchronous teaming expose several unmet coordination requirements. Here are the common challenges that modern communication protocol needs to handle to enable scalability, robustness, and distributed communication among agents:

\subsubsection{Decentralization and Scalability}
Future multi-agent communication should support peer-to-peer, fault-tolerant dissemination of Information. In a large swarm of agents, relying on central brokers or one-to-one messaging does not scale well and introduces single points of failure. Gossip protocols exemplify the needed approach: a decentralized, epidemic-style communication where each agent periodically shares updates with random peers. Over time, this rumor-mongering ensures data (e.g., a discovery or an alert) reaches all agents without a central coordinator. Each message exchange is small and local, but through repeated stochastic interactions, the whole group attains a consistent view. This property is ideal for swarms because it reduces communication bottlenecks and gracefully handles node failures – no single agent is irreplaceable in the information flow \cite{milvus}. Similarly, DMARL \cite{cheruiyot2025survey} and decentralized policy learning to distribute gradients, rewards, and model checkpoints.

\subsubsection{Dynamic Membership and Discovery}
In future agent ecosystems, agents may join or leave the network frequently or form ad-hoc coalitions based on context. Protocols must enable on-the-fly discovery of new agents and integration of their knowledge. Traditional MACPs, e.g., A2A, assume a registry or announcement mechanism for agents to find each other. Gossip will be able to provide a continuous background process of discovery – as new agents begin gossiping, their presence and any state they contribute propagate naturally. This Gossip-based discovery approach avoids the need for complex directory service agents, as the shbor set evolves to incorporate newcomers, and Information about who is in the network spreads organically.

\subsubsection{Minimal Overhead \& Robustness}
Gossip communication is lightweight and resilient. Each agent only contacts a few others at any time, which keeps per-agent overhead low even as the overall system grows large. The redundancy of random exchanges means even if some messages are lost or some agents go offline, the Information will likely still reach others via alternate paths. For mission-critical multi-agent operations, this redundancy is vital for reliability. By focusing on local interactions rather than global broadcasts, Gossip-based protocols self-organize the network around changing conditions \cite{milvus}. In essence, the communication fabric becomes adaptive: Information finds its way through whatever routes are available, much like packets on the internet routing around outages.

\subsubsection{Context Sharing and Emergent Consensus}
In a swarm-intelligent system, agents need to share situational context (observations, partial results, warnings) freely so that a coherent global behavior emerges from local actions. For example, EGC agents \cite{mcDonnell2020evolved} Goss "p "contra "ts" (capabilities + intents) and evolve collaborative strategies via distributed learning. These frameworks enable emergent collaboration without centralized planning. Future protocols should support indirect, asynchronous context pooling – something Gossip is naturally good at. For example, suppose one agent discovers a critical fact (say, a change in the environment or a new user request). In that case, Gossip-style updates can ensure all agents gradually learn this fact~\cite{milvus}. Over a series of Gossip cycles, the swarm reaches consensus or at least a standard knowledge baseline without any master agent telling everyone directly. This stands in contrast to current protocols (MCP, A2A), which excel at directed messaging and structured dialogues but do not inherently implement epidemic dissemination. Incorporating a Gossip layer would enable emergent agreement and collective learning in a group of agents – a key requirement for swarm intelligence. 

\subsubsection{Security and Veracity in Decentralized Exchange}
One challenge is that Gossip protocols, by nature, propagate Information blindly, which could include false or malicious data. Thus, future agent communication will need a hybrid approach for propagation combined with trust mechanisms (perhaps cryptographic signatures or consensus validation steps). An advanced swarm protocol might let agents gossip metadata about information sources and confidence levels, so peers can decide whether to adopt that info. While current protocols like MCP emphasize security and identity (authenticating senders, controlling tool access) \cite{habiba2025emergence}, these measures must extend to a decentralized context. Each agent should assess messages with built-in skepticism, only integrating knowledge that passes integrity checks or is corroborated by multiple peers. Designing a secure "Gossip" mechanism is an open research problem, but it is a need for future autonomous agent swarms. Gossip ensures no centralized broker, aligning with your Federated Coordination and Legal Entity patterns, and also the decentralized nature of the system; therefore,  security and privacy of the system remain intact.

In conclusion, while the A2A protocol lays the foundation for interoperable communication among agents, it lacks orchestration primitives necessary for decentralized, adaptive, and resilient agent ecosystems. Gossip protocols provide a promising augmentation, introducing mechanisms for emergent behavior, dynamic coordination, and fault tolerance. Future work should explore hybrid models where formal message-passing (as enabled by A2A) is paired with informal Gossip layers to achieve more prosperous and more resilient agentic architectures.

\section{Rethinking Agent Communication — Mechanics, Capabilities, and Limits of Gossip Protocols}

\begin{figure}[!htb]
\centering

\includegraphics[width=0.85\linewidth]{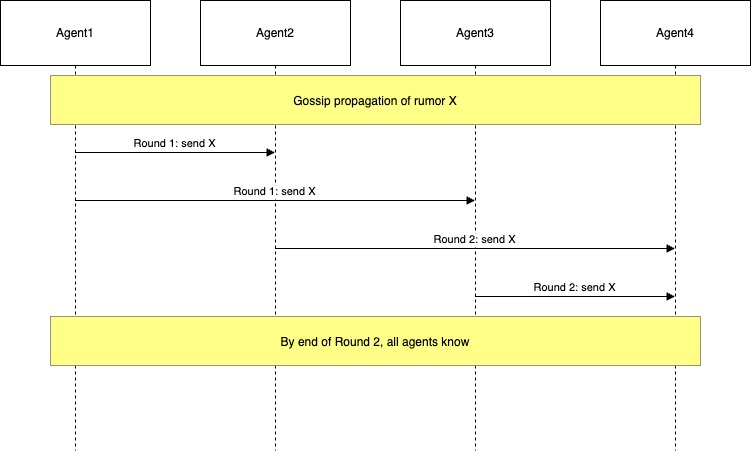}
\caption{ For example, of Gossip protocol dissemination. In Round 1, Agent 1 gossips a new piece of Information X to two peers (Agents 2 and 3). In Round 2, those agents forward X to another peer (Agent 4). After two rounds, all four agents have received X. Gossip exchanges continue periodically, providing redundant paths and eventual consistency.}
\label{fig:gosssip-propagation}
\end{figure}

The Gossip Protocol is a decentralized communication mechanism inspired by how rumors spread in social systems. Gossip protocols originated in foundational distributed systems research as a method for disseminating updates in a manner analogous to disease spread\cite{Montresor2007}. Each node (or agent) periodically exchanges state information with a random neighbor, gradually propagating knowledge through the network
\cite{highscalability.com}. This peer-to-peer epidemic propagation has key properties desirable for multi-agent AI: Each node (or agent) randomly selects peers to exchange Information with at regular intervals. Figure~\ref{fig:gosssip-propagation} shows how agents communicate over the gossip protocol. 

From an analytical perspective, Gossip protocols have been shown to possess robust guarantees. For instance, under random peer selection, the spread of a rumor behaves like a branching process that, with high probability, reaches all nodes rapidly (with exponential convergence)
\cite{Procaccia2007}. 

\subsection{Core Mechanics of Gossip Protocols}

 At its core, a gossip protocol works like an epidemic. Each agent (node) randomly selects a subset of other agents to exchange Information with during periodic gossip rounds\cite{highscalability.com}. Key properties of the gossip protocol include:

 \subsubsection{Distributed Topology}
Gossip protocols establish a \textit{many-to-many} communication pattern that evolves dynamically over time. Each agent engages in periodic, randomized exchanges with peers, leading to emergent system-wide connectivity through chains of local interactions. Unlike MCP or A2A protocols, where communication is typically one-to-one and transaction-bound, gossip maintains ongoing interaction flows. In MCP/A2A, an agent must either explicitly address multiple peers or employ an orchestrator to disseminate information, often leading to tight coupling and communication overhead. In contrast, gossip achieves eventual global dissemination via local redundancy.

A notable challenge in gossip-driven topologies is handling information redundancy, irrelevance, or staleness. While humans intuitively apply semantic filtering to social gossip, autonomous agents require explicit logic to determine what information to discard, forward, or process—posing significant design complexity for agent developers.

\subsubsection{Communication Pattern}
Gossip protocols follow a peer-to-peer, symmetric interaction model. Any agent may initiate exchanges with others, typically chosen via uniform random sampling. For example, an agent might periodically share resource utilization data (e.g., "Agent X at 90\% CPU, low battery") or broadcast capability metadata ("I can analyze images"). This distributed, redundant sharing allows for robust, decentralized load balancing—tasks are implicitly directed toward agents with more available resources, without requiring central control.

Such patterns bolster system fault tolerance. Even if an agent fails mid-broadcast, the same information likely propagates through other paths. This redundancy underpins self-healing behaviors, enabling agents to collectively detect hot spots, adjust behavior, or reroute tasks. In heterogeneous societies where agents have diverse skills (e.g., language translation, image recognition), gossip acts as a decentralized registry: capabilities are advertised continually, allowing agents to gradually develop a partial view of the skill distribution across the network.

\subsubsection{Content and Semantics}
Unlike structured communication protocols (e.g., A2A or MCP), which define explicit message types (e.g., "ActionRequest", "Observation") and ontologies, gossip is content-agnostic. There is no enforced schema or predefined semantic structure. As a result, agents bear the responsibility of interpreting payloads locally.

While this model offers implementation flexibility, it complicates interoperability and safety in heterogeneous systems. For instance, differing agent architectures may encode the same capability differently, risking misinterpretation. By contrast, structured protocols facilitate plug-and-play integration between agents by standardizing semantics. This trade-off—semantic freedom versus coordination reliability—should be evaluated based on the system's scale, heterogeneity, and criticality.

\subsubsection{Emergence vs. Intention}
Gossip protocols are well-suited for emergent, uncoordinated behaviors, such as spontaneous consensus formation or background knowledge sharing. In contrast, structured protocols excel in goal-directed interactions—when Agent A explicitly delegates a sub-task to Agent B with well-defined expectations.

When the origin and destination of information are known a priori, direct A2A messaging is more efficient. However, in systems where any agent might produce or consume data, and where systemic awareness is beneficial (e.g., for maintaining a shared environmental model), gossip excels. It enables ambient coordination without the fragility of orchestrated workflows.

\subsubsection{Failure Detection and Recovery}
Gossip protocols are historically rooted in distributed systems for membership and failure detection—e.g., SWIM [Das et al., 2002] and Fireflies [Johansen et al., 2015]. These principles translate naturally to multi-agent networks, where each agent periodically shares the list of live peers it has recently interacted with. If an agent goes silent for a set number of gossip intervals, it is suspected to have failed. As these suspicions propagate, agents converge on a consistent view of network membership.

Upon rejoining, an agent's reappearance is similarly gossiped about, leading to eventual reintegration. This decentralized failure detection replaces the need for centralized heartbeat servers. Additionally, gossip can propagate intent (e.g., "I am inspecting Location Q") or state changes ("Entering busy mode"), facilitating indirect, distributed coordination. In swarm robotics or multi-UAV settings, such behavior enables dynamic alignment of plans (e.g., avoiding redundant coverage) without a central planner.

\subsubsection{Performance}
Structured protocols provide deterministic, low-latency communication between specified agents. If Agent A knows Agent B holds the needed data, an A2A query is optimal. Gossip, by contrast, involves multiple hops and probabilistic propagation, leading to higher latency for individual requests.

However, gossip shines in one-to-many communication and fault-tolerant dissemination. For example, spreading a new data point to agents via A2A incurs load on the originator; gossip distributes this burden across the network, maintaining messages per round but avoiding central bottlenecks. For small-scale or critical systems, A2A remains favorable; for large-scale, loosely coupled environments, gossip offers superior scalability and resilience.

\subsubsection{Coordination Paradigm}
Gossip fosters a reflexive coordination model based on state-sharing and convergence, rather than explicit negotiation. Agents do not assign roles or tasks directly; instead, they share observations, intentions, and capabilities until a consistent global state emerges to guide decentralized decision-making. This biologically inspired model (e.g., bee dances, termite construction) emphasizes robustness and adaptability over planning precision.

Conversely, A2A and MCP resemble digital coordination, where agents exchange explicit requests, delegate subtasks, and rely on transactional guarantees. This enables complex workflows and task contracts—but at the cost of rigidity and vulnerability to failures.

\begin{figure}[!htb]
\centering
\begin{subfigure}[b]{0.48\textwidth}
\centering
\includegraphics[width=\textwidth]{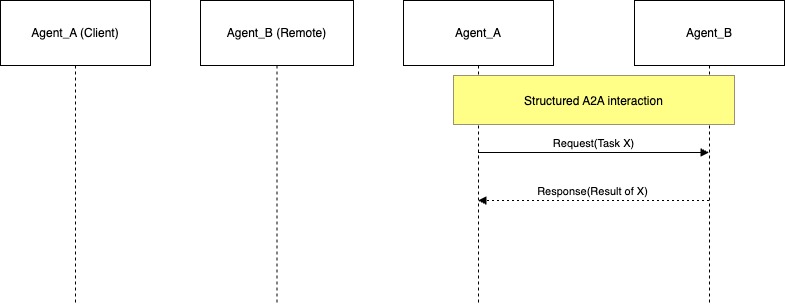}
\caption{Structured A2A Communication}
\label{fig:a2a-communication}
\end{subfigure}
\hfill
\begin{subfigure}[b]{0.48\textwidth}
\centering
\begin{tikzpicture}[
agent/.style={circle, draw, minimum size=1.2cm, align=center},
Gossip/.style={->, thick, green, bend left=15},
network/.style={rectangle, draw, dashed, minimum width=6cm, minimum height=4cm, align=center}
]
\node[network] at (2,0) {Gossip Network};
\node[agent] (agent1) at (0.5,1) {A1};
\node[agent] (agent2) at (3.5,1) {A2};
\node[agent] (agent3) at (3.5,-1) {A3};
\node[agent] (agent4) at (0.5,-1) {A4};
\draw[Gossip] (agent1) to (agent2);
\draw[Gossip] (agent2) to (agent3);
\draw[Gossip] (agent3) to (agent4);
\draw[Gossip] (agent4) to (agent2);
\node[text width=4cm, align=center] at (2,-2.5) {Emergent peer-to-peer propagation};
\end{tikzpicture}
\caption{Gossip Network Communication}
\label{fig:Gossip-network}
\end{subfigure}
\caption{Structured vs. Gossip communication. Left: Direct A2A communication between two agents for targeted task delegation. Right: Gossip-based propagation where state awareness spreads implicitly and redundantly across the network.}
\label{fig:communication-comparison}
\end{figure}
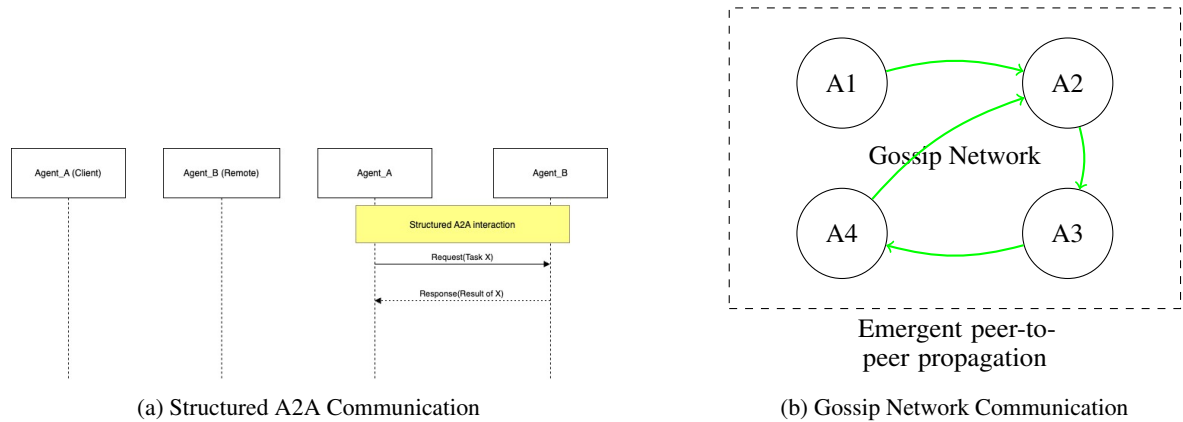

Figure~\ref{fig:communication-comparison} contrasts these two paradigms: direct messaging ensures immediate response, while gossip gradually saturates the network with state information.

\begin{figure}
\centering
\includegraphics[width=0.75\linewidth]{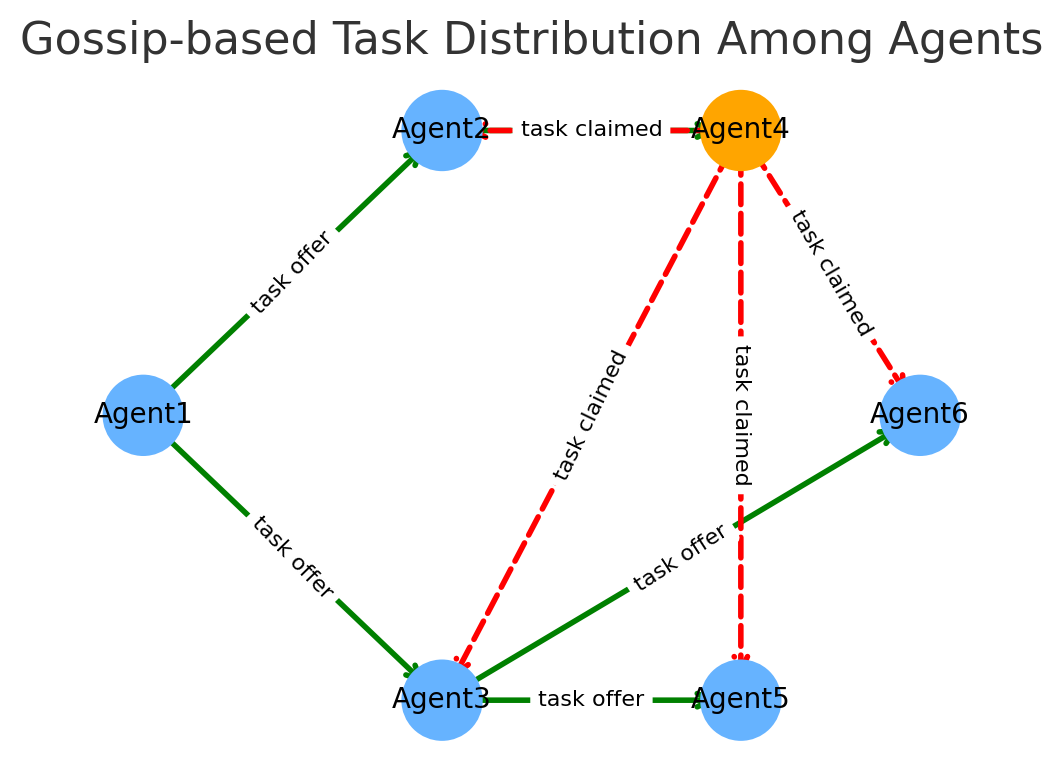}
\caption{Gossip-based task distribution among agents. Green arrows show task availability propagation; red arrows show task claimed updates. This decentralized mechanism reaches eventual consistency without a central coordinator.}
\label{fig:gossip}
\end{figure}

\subsection{Use Cases for Gossip Protocols}

Gossip-based communication is a powerful tool, but it is not a silver bullet for all coordination problems. Let's delineate the boundary conditions where gossip excels:

\subsubsection{Agent/Service Discovery}

In large-scale or highly dynamic networks (think hundreds of agents possibly joining/leaving unpredictably), gossip shines for discovery and membership management. It avoids any single point of failure and can gossip Information about new agents or departing ones quickly across the system. This is useful in ad-hoc networks, IoT swarms, or open multi-agent ecosystems where a fixed registry might not be feasible. Each agent eventually learns who is out there, even if, at any moment, views might be partial. The system remains eventually aware of its constituents.

\subsubsection{Task Pooling and Emergent Coordination}
As described earlier, gossip is particularly useful in scenarios such as task allocation, load balancing, or sensor fusion, where a global picture emerges from many local communications. If the problem can tolerate a bit of asynchrony, gossip allows agents to solve it without heavy infrastructure collectively. For example, in a swarm of drones searching an area, gossip can disseminate which regions have been covered, leading drones to autonomously cover new areas without centralized assignment – a form of emergent coordination. The redundancy provides fault tolerance: even if a few drones crash, their last known Information has already been shared with others who continue the mission.

\subsubsection{Failure Detection and Recover} 
Gossip's robustness and eventual consistency make it ideal for failure detection and dissemination of status in a distributed system. Protocols like SWIM leverage gossip to detect failures in $O(1)$ time on average and inform all nodes of membership changes in $O(log n)$ time. In multi-agent terms, if some agents die or become unreachable (perhaps a server went down or a robot lost power), gossip quickly spreads that knowledge. Hence, others stop depending on the failed agents and possibly reassign tasks. Compared to a centralized heartbeat monitor, gossip-based failure detection has no single point of failure and scales to large clusters easily. The tradeoff is a small chance of false positives and slight delays in detection – but these are often acceptable in tradeoff for high availability.

\subsubsection{Scalable Knowledge Sharing}
Whenever you need to keep a large number of agents roughly in sync on some state (like a cache of data, a shared ledger, or environmental knowledge) without requiring instant consistency, gossip is a top choice. For instance, something like the global trending topics in a social network of agents could be gossiped – all agents eventually learn the trends without a central server. Systems inspired by gossip (e.g., Amazon Dynamo's storage system) manage to replicate data across hundreds of nodes with eventual consistency but extreme fault tolerance – a valuable model for multi-agent data sharing as well. In brief, use gossip when you prioritize availability, partition tolerance, and scalability over immediate consistency.

However, there are also scenarios where the gossip protocol does not work:

\subsubsection{Strict Ordering Requirements:}

If the multi-agent application requires certain events or actions to occur in a globally serialized order (for example, a sequence of financial transactions or game moves that all agents must agree on exactly in order), gossip alone is not suitable. Gossip will deliver updates out of order and at different times to different agents, which could violate a strict consistency condition. For such cases, consensus algorithms (RAFT, Paxos) or centralized sequencing are needed. Gossip can distribute Information about an event, but not in a deterministic order. For example, two agents might gossip conflicting commands to a robot; by the time the conflict is resolved, the robot might have acted on one. If you genuinely need linearizability or total order, Gossis isn't the right tool. (However, gossip can still support those systems in the background, e.g., gossiping about a leader's decisions for redundancy, but the decision itself must come from a coordinated mechanism.).

\subsubsection{Immediate Consistency Needs }

Building on the above, any scenario where all agents must have the exact same view at the same time is ill-suited to gossip. Gossip is eventually consistent – there will be moments when different agents have different views. If, say, an agent making a query needs up-to-date info from all other agents, gossip might be too slow or inconsistent. A concrete example: a distributed lock or a mission-critical boolean flag (launch missile or not) that all agents check – if one agent flips it via gossip, another agent might not see it for a few gossip rounds, which could be disastrous. In such cases, a strongly consistent approach (like a central lock manager or immediate broadcast via reliable multicast) is used. Gossip sacrifices consistency for availability; if you can't sacrifice consistency, you shouldn't rely on gossip alone. Cassandra's designers note that gossip only works when operations are commutative or ordering is not needed.

\subsubsection{Low-Latency, High-Precision Coordination}
If decisions need to be made immediately and with absolute agreement, gossip might be too latent. For example, high-frequency trading agents coordinating a transaction in milliseconds might find gossip rounds (which might be tens or hundreds of milliseconds sluggish, and the uncertainty of who knows what when is problematic. Similarly, suppose an agent needs confirmation that all others received a command before proceeding. In that case, gossip's probabilistic nature is troublesome; there's always a non-zero chance someone hasn't gotten the memo yet. In such cases, a synchronous barrier or acknowledgment protocol is used rather than gossip. Gossip also tends to introduce a lot of "f "no "se" (redundant messages), which can be an issue on constrained networks (e.g., low-bandwidth radio networks might not handle constant gossip traffic well if the group is large).

\subsubsection{Sensitive or Confidential Information} Gossip's strength is wide propagation, which is a weakness if you want to limit information flow. In some multi-agent settings, specific data must stay local or only be shared with specific trusted parties. Gossip, unless augmented with heavy encryption and access controls, is a broadcast flood. Even if it's hard to target only relevant agents via pure gossip (the nature is to propagate randomly), for sensitive environments (e.g., agents handling personal data or secure military info), a direct point-to-point encrypted channel or a broker that enforces access rules might be mandated instead of gossip. Gossip can be made safe, but it introduces an extra layer of complexity, and its default behavior conflicts with the principle of least privilege.

\subsection{Security and Trust Considerations}
While gossip is powerful, its very features (random propagation, lack of central authority, redundancy) introduce challenges in adversarial or noisy environments. Key concerns and possible remedies include:

\subsubsection{Propagation of False or Malicious Data:}
By default, gossip will spread anything – including incorrect or malicious Information – with the same fervor as truth. A rogue agent could inject false rumor" ("Agent X is d "wn" whit'st's not, or a fake task) and gossip would dutifully propagate it to many others. To mitigate this, message authenticity and integrity checks are critical. Agents should cryptographically sign their gossip messages, and include non-repudiable identifiers \cite{johansen2015fireflies}. For example, if the state update is signed, others can verify the origin and detect tampering. A certificate authority or web-of-trust can help establish verifiable identities. In secure gossip systems like Fireflies, each gossip message (like a membership accusation) is accompanied by digital signatures so that malicious nodes cannot forge others' messages. This ensures that even though gossip is peer-to-peer, receivers can filter out Information that the purported sender didn’t validly endorse.

\subsubsection{Amplification and Flooding Attacks} Gossip's redundancy means a small message can multiply through the network. An attacker might exploit this by initiating many bogus updates to flood the system (a form of denial-of-service via gossip). Rate limiting and spam suppression mechanisms can be introduced. For instance, agents could use probabilistic filtering (ignore messages that appear too frequently or from suspicious origins) or require a proof-of-work for injecting new significant updates. Additionally, techniques like randomized peer selection\cite{highscalability.com}and time-to-live on gossip messages can help. By limiting how far or how long a given message can travel (e.g., a gossip TTL field or hop count), the network can dampen endless amplification. Some systems employ gossip admission control – e.g., an agent will only forward at most N distinct new messages per unit time, or will drop messages that don't meet specific validity criteria. Such controls introduce a trade-off (slightly slower propagation) but improve robustness against spammy behavior. 

\subsubsection{Staleness tradeoffict Resolution:}
Because gossip is eventually consistent, there can be windows of inconsistency where different agents have different values for t "e "s "me" Information. For example, two agents might concurrently gossip conflicting updates ($Agent A$ says $Resource Y = 5$ while $Agent B$ says $Resource Y = 7$). Without coordination, these differences could circulate and overwrite each other repeatedly. To handle this, gossip protocols attach version numbers or timestamps to data and implement merge rules. Typically, the rule "s "keep the latest update as determined by logical clocks or timestamps\cite{gossip-performance-cassandra}. Anti-entropy exchanges ensure that when agents gossip, each takes the highest version of each data item from the other \cite{highscalability.com}. This way, conflicts are resolved in favor of one update (usually the newest) and obsolete Information gets overridden. More complex states can be managed via CRDTs (Conflict-Free Replicated Data Types) – data structures designed such that concurrent updates merge naturally without conflict. By using CRDTs for shared state (like counters, sets, etc.), gossip can disseminate state updates from multiple agents in any order and still converge to a consistent result. In essence, CRDTs mathematically guarantee eventual consistency under gossip by making all operations commutative. Systems like Cassandra use simple versions of this idea (each piece of data has a timestamp; the most recent timestamp wins) \cite{gossip-performance-cassandra}. The tombstone mechanism in Cassandra is another example: when data is deleted, a tombstone marker (with a version) is gossip instead of being removed immediately, so that all replicas agree on the deletion before final purge.

\subsubsection{Reputation and Redundant Validation:}
Another strategy against malicious or faulty inputs is to incorporate reputation or multi-source validation into gossip. Instead of unquestioningly trusting the first message, an agent can wait until it hears the same fact from multiple independent peers before acting on it. If one agent gossips that Agent Z failed, a second opinion from a different neighbor increases confidence. This can be formalized by requiring k-confirmations – an agent treats a gossip update as credible only after hearing it k times from other peers. Honest gossip ensures that actual events will be reported by many, whereas a false rumor from a single bad actor might not get corroborated (assuming not too many colluding bad actors). Over time, agents could also maintain a reputation score for peers based on past behavior. If specific agents often send gossip that contradicts the majority or is later proven false, their messages might be discounted or ignored (low reputation). Gossip networks can thus evolve a trust overlay on top of the raw protocol. This is analogous to human gossip: Information coming from multiple trusted friends is believed more than from a stranger. Technically, implementations might piggyback trust metadata (like scores or blocklists) in gossip messages, so agents collectively learn who is reliable. An example in literature is gossip-based reputation systems for swarms, where agents gossip both about the environment and about each other's trustworthiness \cite{johansen2015fireflies}. Secure gossip frameworks like Fireflies also address Sybil attacks (where one entity creates many fake nodes): by using a combination of certificates and limiting the fraction of newcomers that can join, they ensure a bounded probability of any gossip peer set having too many compromised members \cite{johansen2015fireflies}

In summary, adding security to gossip involves authentication, validation, and rate control. These measures combat the open-ended nature of gossip, allowing the benefits of redundancy and decentralization to be enjoyed without succumbing to malicious exploits or chaotic data.

Gossip is best viewed not as a replacement for structured communication, but as a complementary layer enabling resilience, discovery, and emergence. For agentic systems operating at scale under uncertainty, it provides essential primitives. Yet for deterministic coordination, trust enforcement, and structured collaboration, additional mechanisms are necessary. Future work should explore hybrid architectures that dynamically combine the strengths of both.

\section{Research Agenda: Key Open Questions}

While gossip protocols have demonstrated utility in distributed systems, their adaptation to agentic AI environments introduces a set of novel and unresolved challenges. These challenges span semantic representation, communication learning, trust calibration, and evaluation methodology. This section outlines concrete research questions and corresponding subdomains that must be addressed to elevate gossip protocols to first-class primitives in agentic multi-agent architectures.

\subsection{Semantic Filtering and Message Compression}

Traditional gossip mechanisms assume flat or scalar state dissemination. However, modern agents operate over high-dimensional, symbolic, and semantically rich state representations (e.g., contextual embeddings, planning intents, or tool trajectories). Naïvely gossiping such information is bandwidth-intensive, semantically redundant, and potentially misaligned with peer relevance.

\textbf{Research Question:} How can agents perform learned compression or semantic abstraction prior to gossip dissemination?

\textbf{Approach:}
\begin{itemize}
  \item Develop message summarization models (e.g., transformer-based compressors) that map agent state to a lower-dimensional semantic representation.
  \item Define a task-conditioned filtering policy $\pi_{f}: S \rightarrow \{0, 1\}$ that determines whether a state fragment is gossiped, based on relevance to current group objectives.
  \item Investigate loss functions for maximizing downstream task utility given lossy gossiped information (e.g., multi-agent intent matching or shared context alignment).
\end{itemize}

\subsection{Trust and Veracity in Open Communication Graphs}

Gossip’s probabilistic and redundant dissemination makes it vulnerable to unverified, malicious, or stale information spread. In heterogeneous agent populations (e.g., multi-vendor or federated settings), trust cannot be assumed.

\textbf{Research Question:} How can trust metrics be incorporated into gossip propagation and influence the credibility-weighted state update?

\textbf{Approach:}
\begin{itemize}
  \item Define a dynamic trust function $T_{i,j}(t)$ between agents $i$ and $j$ to govern message weighting.
  \item Integrate cryptographic provenance (e.g., Merkle proofs or source-chained digests) into gossip payloads.
  \item Apply techniques from reputation systems and Bayesian belief networks to build agent-specific trust priors that evolve with communication history.
\end{itemize}

\subsection{Learning Adaptive Gossip Policies}

Current gossip implementations are static and protocol-driven. In contrast, agentic communication environments demand context-aware, bandwidth-optimized, and goal-aligned dissemination.

\textbf{Research Question:} Can agents learn dynamic communication policies that optimize what, when, and whom to gossip with, given environmental and task context?

\textbf{Approach:}
\begin{itemize}
  \item Formulate a reinforcement learning setting where each agent learns a communication policy $\pi_{\text{gossip}}: \mathcal{S} \rightarrow \mathcal{M}$ mapping state to message set and peer selection.
  \item Introduce a reward signal based on group-level convergence metrics (e.g., consensus quality, coordination entropy) or task performance.
  \item Extend MARL frameworks (e.g., QMIX, MAPPO) to include communication bandwidth as a cost term in the reward function.
\end{itemize}

\subsection{Temporal Consistency and Staleness Bounding}

In dynamic environments, the delay introduced by gossip propagation can lead to action misalignment due to stale context. Unlike synchronous protocols, gossip offers no temporal guarantees.

\textbf{Research Question:} How can temporal relevance be preserved or inferred in asynchronously gossiped state information?

\textbf{Approach:}
\begin{itemize}
  \item Integrate temporal metadata such as \textit{age-of-information} (AoI) and expiration TTLs into each gossiped packet.
  \item Design aging-aware update rules that discount old information in belief update functions.
  \item Model gossip state as a time-decayed Markov random field and analyze convergence properties under bounded delay distributions.
\end{itemize}

\subsection{Robustness Under Churn and Partial Connectivity}

Agentic systems deployed in real-world environments (e.g., robotics swarms, disaster response) must remain robust under node failures, join/leave events, and intermittent connectivity.

\textbf{Research Question:} What are the theoretical limits and empirical failure thresholds of gossip convergence in non-stationary agent topologies?

\textbf{Approach:}
\begin{itemize}
  \item Model agent churn as a stochastic process and simulate gossip resilience under varying dropout rates and reconnection delays.
  \item Evaluate convergence probability and delay under different peer selection strategies (uniform vs degree-biased gossip).
  \item Analyze information coverage over dynamic Erdős–Rényi or scale-free communication graphs using probabilistic graph theory.
\end{itemize}

\subsection{Emergent Coordination via Gossip Dynamics}

Beyond state sharing, gossip may serve as an implicit substrate for emergent planning, task negotiation, or shared policy convergence—without requiring centralized planning.

\textbf{Research Question:} Can gossip protocols act as low-level communication substrates enabling emergent alignment or distributed decision-making?

\textbf{Approach:}
\begin{itemize}
  \item Investigate gossip-driven convergence in goal vectors, local policies, or shared belief spaces under decentralized reinforcement learning.
  \item Define gossip-based consensus operators over symbolic structures (e.g., partial plans, tool graphs) and evaluate alignment dynamics.
  \item Extend collective intelligence models (e.g., flocking, Boids, stigmergy) with gossip-augmented state propagation.
\end{itemize}

\subsection{Benchmarking and Evaluation Methodology}

Despite increasing interest in agent communication, no standardized benchmark exists to evaluate gossip-based coordination in task-oriented environments.

\textbf{Research Question:} What metrics and environments best measure the effectiveness and scalability of gossip-enhanced multi-agent systems?

\textbf{Approach:}
\begin{itemize}
  \item Propose benchmark environments (e.g., decentralized search-and-rescue, swarm logistics, decentralized toolchain invocation) where gossip plays a central role.
  \item Define metrics such as \textit{semantic consensus entropy}, \textit{convergence time}, \textit{network load per node}, and \textit{resilience to message loss}.
  \item Compare hybrid (gossip + structured) communication stacks against pure MCP/A2A baselines in simulation.
\end{itemize}

\bigskip
In addressing these challenges, we open a research pathway toward communication frameworks that support not only secure task execution but also emergent, adaptive collective intelligence. Rather than replacing structured protocols, gossip may serve as the missing substrate for reflexivity, resilience, and distributed cognition in agentic AI systems.

\section{Use Cases: Agent-Centric Applications of Gossip Protocols}

\subsection{Industrial Automation: Coordination Among Autonomous Agents}

In modern smart factories, autonomous agents — including robotic arms, mobile conveyors, and inspection bots — operate collaboratively to manage assembly tasks. Rather than relying on centralized orchestrators, future manufacturing systems envision each machine as an intelligent agent capable of local decision-making and peer-to-peer coordination. Gossip protocols provide a scalable substrate for these agents to share state, infer collective behavior, and adapt dynamically.

\textbf{Context Sharing for Production Awareness:} Each agent periodically gossips its current state (e.g., task queue, workload, material inventory, or maintenance alerts) to a few peers. Over successive rounds, this enables all agents to converge on a shared understanding of the global production context. For instance, if one robotic welder slows due to heat buildup, nearby feeders can delay inputs accordingly — without centralized intervention. This context-sharing supports swarm-level adaptation and avoids local bottlenecks.

\subsubsection{Decentralized Task Allocation}

In agent-based factories, task assignment can emerge through gossip. Suppose a new job is detected (e.g., a high-priority assembly). The discovering agent gossips a "task available" message. Peer agents evaluate their current load, capabilities, and relevance, and those suited may respond — either via direct messages or by amplifying interest signals through gossip. Over a few rounds, the task naturally migrates to an available and capable agent. This model supports emergent auctions or self-organized negotiations without needing predefined leaders or task planners.

\subsubsection{Swarm-Level Load Balancing}

To maintain consistent throughput, agents must balance workloads dynamically. Gossip-based averaging allows agents to estimate system-wide production metrics (e.g., average latency or defect rates) and modulate their own behavior accordingly. If one agent lags, its peers may adjust cycle times to compensate. This emergent equilibrium mimics cruise control in coordinated systems — but arises from local information and gossip exchanges alone.

\subsubsection{Fault Detection and Redundancy Awareness}

Instead of relying on external monitors, agents can gossip heartbeat signals or behavioral anomalies. When an agent fails or becomes inconsistent, the lack of expected updates is detected and propagated. Peers may reroute material flow, reassign pending tasks, or trigger redundant capabilities. Through gossip, fault awareness spreads rapidly and robustly — and agents collectively reconfigure without top-down commands.

\textbf{Agentic Challenges:} Real-time synchronization (e.g., precise part handoffs) may require deterministic timing beyond gossip’s scope. In such cases, gossip handles macro-level coordination (task distribution, state convergence), while micro-level timing is managed by local policies. Another challenge is message relevance: agents must prioritize which state fragments are worth gossiping to avoid overload — opening opportunities for learned gossip policies or semantic compression techniques.

\subsection{Disaster Response: Self-Directed Coordination in Uncertain Environments}

In disaster scenarios — earthquakes, floods, or wildfires — agents such as drones, autonomous vehicles, and ground robots are deployed to navigate dynamic, degraded environments. Infrastructure is unreliable, and tasks evolve unpredictably. Agents must operate with partial knowledge, collaborate ad hoc, and adapt continuously. Gossip-based communication is well-suited to this class of decentralized, reflexive systems.

\subsubsection{Peer-Driven Information Diffusion}

Agents gossip newly observed facts (e.g., blocked roads, victim sightings, structural damage) to neighbors. Each agent maintains a buffer of local observations and shares it opportunistically. Over time, the network converges on a common understanding of the terrain and mission priorities. This model requires no global clock, centralized planner, or perfect connectivity. Mobile agents act as data ferries — ensuring slow but reliable convergence even across partitions.

\subsubsection{Reflexive Task Coordination}

Instead of being assigned roles, agents gossip about their actions and intentions ("searching Building 3", "scanning Zone A"). Others use this implicit plan distribution to avoid duplication and detect gaps in coverage. If no gossip mentions Zone C for several rounds, an agent may explore it. If many agents are clustered, some may relocate. This leads to distributed division of labor driven by emergent consensus rather than fixed roles.

\subsubsection{Dynamic Resilience and Recovery}

If an agent is lost (e.g., crashes or loses power), its absence becomes apparent through lack of updates. Others can infer failure and assume its tasks. Gossip-based recovery requires no pre-registration of backup roles. Agents monitor each other implicitly through updates and fill in gaps as needed — preserving overall mission continuity.

\subsubsection{Message Prioritization and Selective Sharing}

Not all observations are equal. Agents must learn to prioritize: a gas leak alert must propagate faster than an environmental temperature sample. Gossip systems can incorporate message-level metadata (e.g., priority, TTL, confidence) and let agents tune their fan-out and resend logic. For example, life-critical observations may be forwarded in every round; routine ones may be probabilistically suppressed.

\textbf{Agentic Challenges:} Disaster response is a bandwidth- and energy-constrained setting. Agents cannot gossip full sensor logs; instead, they must semantically compress observations or use learned filters. Gossiping stale or misleading information (e.g., a false safe zone) could be dangerous. Agents must incorporate consistency checks, possibly using local agreement (e.g., require confirmation from two sources). Gossip policies must balance urgency, veracity, and propagation cost.

\subsection{Agent-Centric Summary}

In both structured (industrial) and unstructured (disaster) domains, gossip enables agents to act more autonomously, adapt to local conditions, and collectively coordinate without explicit instructions. It supports:

\begin{itemize}
    \item \textbf{Decentralized perception:} agents share and converge on world state via local views.
    \item \textbf{Emergent planning:} tasks are adopted, negotiated, and rebalanced through peer interactions.
    \item \textbf{Collective reflexes:} recovery, rerouting, and reconfiguration happen bottom-up, without hierarchy.
\end{itemize}

Rather than serving as a replacement for structured protocols like MCP and A2A, gossip enriches them by providing the ambient, asynchronous substrate that supports **reflexive, scalable agent collectives**. The future of agentic AI requires this duality: structured reasoning for secure goal-driven tasks, and gossip-based overlays for emergent behavior and swarm-level resilience.

\section{Misconceptions and Research Challenges for Gossip Protocols in Agentic AI}

Despite the longstanding use of gossip (epidemic) protocols in distributed systems such as Cassandra, Dynamo, and Consul, their application in multi-agent, learning-driven, and semantic AI systems remains poorly understood. In this section, we (1) clarify misconceptions that often limit gossip's adoption in agentic workflows, and (2) formulate concrete, forward-looking research challenges to guide its development.

\subsection{Misconceptions About Gossip in Agentic Systems}

\subsubsection{Myth 1: Gossip is Too Slow for AI Coordination.} 
While gossip is probabilistic and periodic, it scales logarithmically with the number of nodes. For example, with a fan-out of 3, a 25,000-agent system converges in ~15 rounds \cite{gossip-performance-cassandra}. At a 1-second interval, this implies complete propagation in under 20 seconds—acceptable for many multi-agent tasks that do not require hard real-time response. Moreover, adaptive gossiping intervals and fan-out adjustment can trade off bandwidth and freshness.
\textbf{Key Insight:} Gossip trades immediacy for fault tolerance and scalability—an acceptable compromise for many agentic use cases such as swarm consensus or ambient context sharing.

\subsubsection{Myth 2: Gossip Leads to Uncontrolled or Inconsistent Behavior.} 
While gossip lacks the determinism of synchronous protocols, modern gossip implementations (e.g., Dynamo \cite{decan2007dynamo}) incorporate versioning, TTLs, and redundancy that ensure eventual consistency. Critically, "inconsistency" in gossip is transient and bounded by probabilistic guarantees—especially when agents use CRDTs or semantic version vectors.
\textbf{Key Insight:} Gossip does not eliminate consistency—it relaxes it to the eventual domain and offers robustness over precision.

\subsubsection{Myth 3: Gossip Is for Infrastructure, Not Intelligent Agents.}
This misconception stems from the dominance of gossip in systems engineering. However, gossip-based learning is gaining traction in multi-agent RL (e.g., GALA \cite{assran2019gala}), actor-critic gossip \cite{springerlearninggossip2024}, and decentralized bandit optimization \cite{chawla2023gossip}. These systems show that gossip can serve as a low-bandwidth substrate for asynchronous knowledge fusion.
\textbf{Key Insight:} Gossip is already migrating from network protocols to learning systems—its use in intelligent agent ecosystems is emerging, not hypothetical.

\begin{table}[!ht]
\centering
\begin{tabularx}{\textwidth}{@{}lX@{}}
\toprule
\textbf{Misconception} & \textbf{Clarification} \\
\midrule
Gossip converges too slowly & Gossip scaling is $O(\log N)$, reaching all nodes within seconds in large systems. Adaptive intervals and targeted fan-out further improve latency. \\
Causes inconsistent state & Gossip achieves eventual consistency using versioning, time-to-live (TTL), and redundant paths. It tolerates node failure better than synchronous protocols. \\
Not used in AI systems & Gossip is increasingly used in MARL, actor-critic learning, decentralized federated learning, and policy-sharing architectures like GALA. \\
\bottomrule
\end{tabularx}
\caption{Reframing common misconceptions about gossip in AI coordination}
\label{tab:gossip-misconceptions}
\end{table}
\subsection{Open Research Challenges in Agentic Gossip Communication}

As gossip protocols are positioned for deployment in cognitively capable multi-agent systems, several technical and theoretical gaps emerge. This section defines a series of open challenges that together form a research agenda for future work on gossip-enhanced agentic architectures. Each challenge is framed as a problem statement with a concrete research hypothesis and potential directions for resolution.

\subsubsection{Challenge 1: Semantic Filtering and Compression}
\textit{Problem:} Gossip protocols traditionally treat all state data uniformly, leading to inefficiencies and semantic noise in large agent networks.
\textit{Hypothesis:} High-dimensional semantic state representations can be learned and compressed into bounded gossip payloads that preserve agent decision fidelity.
\textit{Research Direction:} Investigate embedding-based state encoders, topic-tagged gossip, or adaptive attention mechanisms that modulate gossip content based on task relevance or communication budget \cite{frontiersdisaster}.

\subsubsection{Challenge 2: Trust and Malicious Participation}
\textit{Problem:} Gossip assumes a largely cooperative environment, but real-world multi-agent systems may include malicious or faulty agents.
\textit{Hypothesis:} Distributed trust metrics and local credibility propagation can limit misinformation in open agent networks.
\textit{Research Direction:} Explore lightweight cryptographic signatures, decentralized reputation diffusion, and anomaly-aware gossip filters for adversarial resilience \cite{ijcai2023trust}.

\subsubsection{Challenge 3: Reflexivity and Temporal Validity}
\textit{Problem:} Gossiped messages may arrive late, be redundant, or represent stale observations that reduce coordination quality.
\textit{Hypothesis:} Agents can learn policies to modulate reaction to gossip based on information age, source, and contextual relevance.
\textit{Research Direction:} Develop temporal decay models (e.g., AoI), TTL-adjusted fan-outs, and semantic deduplication layers \cite{ieeedisastergossip2024}.

\subsubsection{Challenge 4: Learning-Integrated Gossip}
\textit{Problem:} Integration of gossip communication into differentiable learning pipelines (e.g., actor-critic, transformer-based agents) remains underexplored.
\textit{Hypothesis:} Gossip can replace or augment centralized gradient sharing with asynchronous, low-bandwidth state synchronization.
\textit{Research Direction:} Extend models like GALA \cite{assran2019gala} to support hybrid LLM-gossip agents that jointly learn what to share, when to share, and with whom.

\subsubsection{Challenge 5: Benchmarking and Evaluation}
\textit{Problem:} There is no standard framework for evaluating the impact of gossip in agentic systems under different environments and workloads.
\textit{Hypothesis:} Gossip-aware agent performance can be assessed using multi-metric environments that capture both task outcomes and coordination quality.
\textit{Research Direction:} Design benchmark tasks using environments such as PettingZoo or Habitat that include metrics like convergence time, consensus accuracy, message load, and recovery from churn or deception.

These challenges illustrate the multi-disciplinary nature of making gossip a first-class coordination substrate in AI. Success will require merging distributed systems, MARL, decentralized trust, and semantic representation learning into cohesive frameworks tailored for agentic reflexivity and collaboration.

\section{Conclusion and Future Work}

As multi-agent systems evolve toward increasingly autonomous and reflexive behavior, their communication protocols must go beyond deterministic orchestration. While modern standards like MCP and A2A provide secure, structured, and verifiable messaging for tool invocation and task delegation, they are fundamentally limited in supporting emergent coordination, decentralized cognition, and dynamic adaptation—capabilities essential for truly intelligent agent collectives operating in open and uncertain environments.

Gossip protocols offer a powerful yet underexplored complement. Rooted in probabilistic dissemination and local interaction, gossip enables scalable, fault-tolerant communication and emergent knowledge convergence without central control. In both predictable domains such as industrial automation and volatile environments like disaster response, gossip has the potential to drive agent self-organization, role redistribution, and real-time context propagation at scale.

However, gossip is not without pitfalls. It lacks semantic awareness, verifiability, and prioritization—making naive implementations brittle in high-stakes contexts. Challenges such as message flooding, stale state propagation, inconsistent intent resolution, and lack of auditability remain open problems. As such, gossip should not be seen as a replacement for structured agent communication, but as a substrate that fills the reflexive gaps MCP and A2A do not address.

This paper calls for a rethinking of multi-agent communication stacks. Rather than choosing between centralized orchestration and decentralized emergence, future systems must integrate both. We envision a hybrid architecture where gossip serves as the ambient coordination layer—spreading soft state, intent signals, and situational awareness—while structured protocols remain responsible for deterministic actions, verifiable negotiations, and security-governed tool access.

To move toward this future, we identify several urgent research directions:

\begin{itemize}
    \item \textbf{Semantic Filtering and Compression:} Develop encoding strategies to distill high-dimensional agent state into gossiped abstractions that retain relevance while minimizing bandwidth and noise.
    \item \textbf{Trust-Aware Propagation:} Embed verification mechanisms—such as digital signatures, peer scoring, or cryptographic attestations—into gossip layers to prevent misinformation, adversarial spread, or Sybil attacks.
    \item \textbf{Reflexive Workflow Assembly:} Design agent architectures in which gossip-triggered context shifts dynamically alter workflows, activate capabilities, or reassign agent roles on-the-fly.
    \item \textbf{Structured–Emergent Bridging:} Create interfaces between gossip-spread information and structured protocol invocation, enabling agents to translate ambient awareness into secure, auditable actions.
    \item \textbf{Benchmarking and Simulation:} Build agent-based benchmarks in both constrained (e.g., smart factories) and unconstrained (e.g., disaster zones) settings to quantify the trade-offs in convergence time, resilience, interpretability, and coordination effectiveness.
\end{itemize}

In conclusion, gossip protocols offer not just redundancy and resilience, but a missing cognitive layer in agentic systems—enabling agents to think, adapt, and align collectively, without central planners or explicit commands. The path forward lies in **engineering hybrid communication fabrics**, where structured intent meets emergent behavior, and where agent intelligence is not only programmed—but collectively grown. We invite the research community to treat gossip not as an old idea, but as a vital thread in the design of future agent ecosystems.

\bibliographystyle{unsrt}
\bibliography{references} 

\end{document}